\documentclass[11pt]{article}
\usepackage{graphicx}
\usepackage{changebar}

\newcommand{\BABARPubYear}    {04}

\newcommand{\BABARConfNumber} {32}
\newcommand{\SLACPubNumber} {10643}

\input babarsym
\def\mumu{$\mu^+\mu^-$}

\def\sphi0{\sin\phi_0}
\def\cphi0{\cos\phi_0}
\def\ra{\rightarrow}
\def\d3ks{D^0 \rightarrow K^0_sK^0_sK^0_s}
\def\b3ks{B^0 \rightarrow K^0_sK^0_sK^0_s}

\def\k2pi{K^0_s\rightarrow\pi^+ \pi^-}
\def\dmpp{\Delta m_{\pi^+\pi^-}}
\def\ak0{\overline{K^0}}
\def\ab0{\overline{B^0}}
\def\bab{B\overline{B}}

\long\def\inst#1{\par\nobreak\kern 4pt\nobreak
    {\it #1}\par\vskip 10pt plus 3pt minus 3pt}

\begin{document}
{\pagestyle{empty}

\begin{flushright}
\babar-CONF-\BABARPubYear/\BABARConfNumber \\
SLAC-PUB-\SLACPubNumber \\
August 2004 \\
\end{flushright}

\par\vskip 5cm

\begin{center}
\Large \bf Measurement of the $\b3ks$ Branching Fraction
\end{center}
\bigskip

\begin{center}
\large The \babar\ Collaboration\\
\mbox{ }\\
\today
\end{center}
\bigskip \bigskip

\begin{center}
\large \bf Abstract
\end{center}
We report a preliminary measurement of the branching fraction for the decay $\b3ks ,$
where the $K^0_s$ mesons are reconstructed through the decay $\k2pi .$
The measurement was performed on a sample of $211\times 10^6$ 
$B\overline{B}$ pairs collected by the \babar\ detector
running on the $\Upsilon (4S)$ resonance at the \pep2\ storage ring.
The branching fraction is measured to be
$${\cal B}(\b3ks )=(6.5\pm 0.8\pm 0.8 ) \times 10^{-6},$$
where the errors are statistical and systematic, respectively.

\vfill
\begin{center}

Submitted to the 32$^{\rm nd}$ International Conference on High-Energy Physics, ICHEP 04,\\
16 August---22 August 2004, Beijing, China

\end{center}

\vspace{1.0cm}
\begin{center}
{\em Stanford Linear Accelerator Center, Stanford University, 
Stanford, CA 94309} \\ \vspace{0.1cm}\hrule\vspace{0.1cm}
Work supported in part by Department of Energy contract DE-AC03-76SF00515.
\end{center}

\newpage
} 

\begin{center}
\small

The \babar\ Collaboration,
\bigskip

%
B.~Aubert,
R.~Barate,
D.~Boutigny,
F.~Couderc,
J.-M.~Gaillard,
A.~Hicheur,
Y.~Karyotakis,
J.~P.~Lees,
V.~Tisserand,
A.~Zghiche
\inst{Laboratoire de Physique des Particules, F-74941 Annecy-le-Vieux, France }
% ...

\end{center}\newpage

\section{INTRODUCTION}
\label{sec:Introduction}
\begin{figure}[!htb]
\begin{center}
\includegraphics[height=4cm]{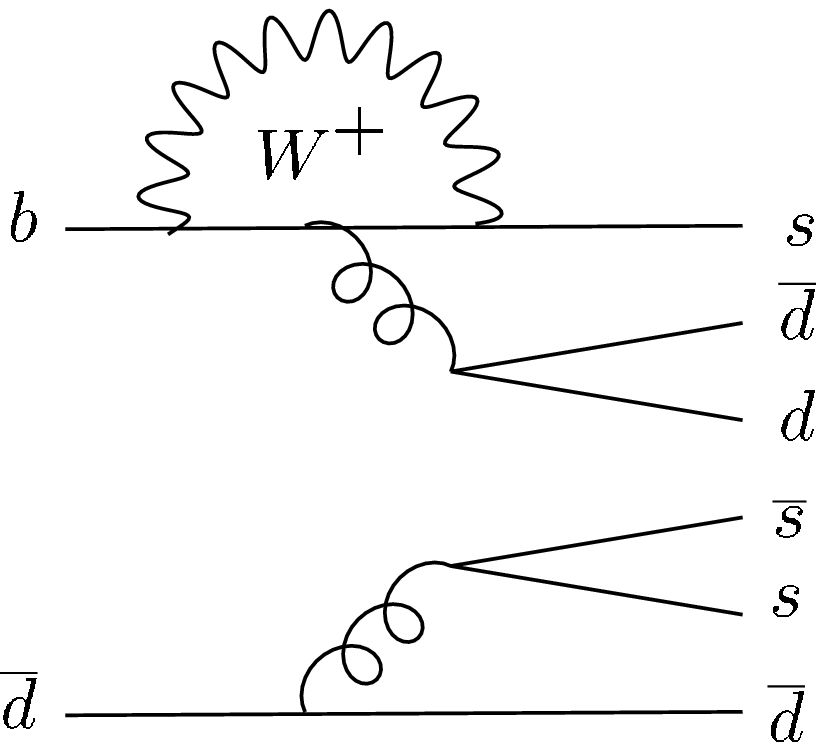}
\caption{The dominant diagram for the decay $\b3ks .$}
\label{fig:b0d0feyn}
\end{center}
\end{figure}

In this paper we report a measurement of the branching fraction $({\cal B})$ for $\b3ks .$
This decay is expected to be penguin dominated; the simplest
diagram that can be drawn without rescattering is shown in Fig.~\ref{fig:b0d0feyn}.
$\b3ks $ is $not$ Cabibbo-suppressed, and so is expected to
have substantially larger branching fraction 
than $B^0\rightarrow 2K^0_s .$

In Ref.~\cite{gronau} the branching fraction ${\cal B}(\b3ks )$ is 
related to ${\cal B}(B^+\ra K^+K^-K^+).$ 
With the assumption of gluonic penguin dominance 
and the usual 
assumption ${\cal B}(K^0\ra K^0_s)={\cal B}(\overline{K}^0\ra K^0_s)=0.5,$
they derive
${\cal B}(B^+\ra K^+K^-K^+)={\cal B}(B^0\ra K^0K^0\overline{K}^0)=
8{\cal B}(\b3ks ).$ 
Using the \babar\ and Belle averaged value \cite{hfag} of
${\cal B}(B^+\ra K^+K^-K^+)=(29.5\pm 1.8)\times 10^{-6},$ 
${\cal B}(\b3ks )$ is expected to be $\sim 4\times 10^{-6}.$

The only previous measurement \cite{belleprd} of this branching fraction
is from the Belle Collaboration, ${\cal B}(\b3ks )=(4.2^{+1.6}_{-1.3}\pm 0.8)\times 10^{-6},$
based on 78 ${\rm fb}^{-1}$ of on-resonance data. The data sample used in this paper is
2.7 the size of that data sample, and
the $\b3ks$ efficiency we estimate for our analysis is larger
than that in Ref.~\cite{belleprd}. 

In this paper we report an inclusive measurement of $\b3ks .$
In addition to non-resonant three-body 
$b\ra s\overline{q}q,$ $(\overline{q}q = \overline{s}s$ or $\overline{d}d)$ 
gluonic penguin decays, 
charmless resonant intermediate states like $B^0\ra f_0K^0$ can produce the $3K^0_s$ final state.
There may also be $b\ra c\overline{c}s$ decays that lead to the $3K^0_s$ final state. 
The dominant of these is expected to be $B^0\ra \chi_{c0}K^0\ra 3K^0_s,$ 
but its product branching fraction \cite{ref:pdg2002}
is $<0.5\times 10^{-6}$ (90\% CL), about a factor of ten smaller
than that expected for $\b3ks .$ 
We do not exclude these from this measurement,
though we will do a 
search for $B^0\ra \chi_{c0}K^0_s$ as a systematic check.  
The product branching fraction for $B^0\ra \overline{D}^0K^0\ra 3K^0_s$ is estimated \cite{ref:pdg2002} 
to be $\sim 9\times 10^{-9}$ and therefore will be ignored.

\section{THE \babar\ DETECTOR AND DATASET}
\label{sec:babar}
The data used in this analysis were collected with the \babar\ detector
at the \pep2\ storage ring. We use 191 ${\rm fb}^{-1}$ of data taken at
the center-of-mass (CM) energy of the $\Upsilon (4S)$ resonance (the on-resonance
data sample). These data correspond to $211\times 10^6\ B\overline{B}$ pairs.   

The \babar\ detector is described elsewhere~\cite{ref:babar}.
The important parts of the detector for this analysis are the charged
particle tracking detectors. These consist of five layers of double-sided
silicon-strip detectors between the beampipe and a 40-layer cylindrical
drift chamber, with both axial and small-stereo-angle superlayers. Both
detectors are in a 1.5 T solenoidal magnetic field, and provide excellent
pattern recognition and momentum measurement for reconstruction of
$\k2pi$ decays. The electromagnetic
calorimeter also contributes to this analysis through the reconstruction
of neutral particles which are used along with charged tracks not coming from the
candidate $\b3ks$ decay to form continuum rejection variables.

Large samples of Monte Carlo (MC) simulated events are used throughout this analysis,
to estimate the reconstruction efficiency and to derive parameters used to
describe the signal and background (BG) shapes in the fit for the signal yield.
Except for some parametrized MC samples used to validate the fit, all the MC samples
were generated with GEANT4~\cite{geant}, with the full detector-response simulation,
and reconstructed with the same programs used for data reconstruction.

\section{CANDIDATE SELECTION}
\label{sec:Analysis}

\subsection{$K^0_s$ RECONSTRUCTION}

All $K^0_s$ candidates used in this analysis are reconstructed through the decay $\k2pi .$
Every pair of oppositely-charged tracks that pass a very loose mass selection is
fitted to a vertex. 
All pairs that pass a cut on the vertex-fit mass $(\dmpp =|m_{\pi^+\pi^-}-m_{K^0_s}|<10.8$ MeV)
and a very loose vertex-fit $\chi^2$ probability cut $(P(\chi^2)>10^{-6})$ are retained
for further consideration. 
Several more cuts are applied to the $K^0_s$ candidates
to reject combinatorial background and $K^0_s$ mesons that are not $B$ decay products.
We require the transverse decay-length $r_{DEC}=\sqrt{(x_{K^0_s}-x_{BS})^2+(y_{K^0_s}-y_{BS})^2}$
be between 0.2 and 40.0 cm. 
Here $K^0_s$ refers to the fitted vertex position and $BS$ refers to the
position of the beamspot (the center of the luminous region), 
which is determined in \babar\ approximately every ten minutes. 
The inner $r_{DEC}$ requirement removes random $\pi^+\pi^-$ combinations that most likely come 
from a common point (the event primary vertex or a short-lived secondary
decay) and just happen to have a mass inside the allowed $\dmpp$ range. 
Candidates that fail
the outer $r_{DEC}$ criteria tend to be from calorimeter splash-back tracks that once again just
happen to pass the $\dmpp$ cut. 
We also require that the angle between the transverse flight vector 
(of which $r_{DEC}$ is the magnitude) and the transverse momentum vector 
of the $K^0_s$ be less than $200$ mrad. 
These cuts, along with the continuum rejection criteria described below, 
have been optimized in a signal-blind study to produce the largest significance for
a branching fraction measurement. 
The cuts were optimized using signal MC to represent the $\b3ks$ decay, 
and on-resonance data in mass sidebands, designed to exclude all
signal candidates, to model backgrounds in the signal $m(K^0_s K^0_s K^0_s)$ region. 

\subsection{$B^0$ RECONSTRUCTION}

All combinations of three $\k2pi$ candidates, where none of the candidates share a charged
track, are used to form $\b3ks$ candidates. 
The $K^0_s$ momentum is calculated in the $\pi^+\pi^-$ vertex fit,
but because $m_{\pi^+\pi^-}$ is not constrained to $m_{K^0_s}$ in the vertex fit, we use
$E_{K^0_s}=\sqrt{{\vec{p}}^{\,2}_{\pi^+\pi^-}+m_{K^0_s}^2}$ in place of the $K^0_s$ energy.

We use two kinematic variables to separate the $\b3ks$ signal from backgrounds.
The energy difference $\Delta E = E_B - \sqrt{s}/2$ is reconstructed 
from the energy of the $B$ candidate in the \epem\ CM frame $E_B$ 
and the total energy $\sqrt{s}.$
The  \DeltaE\  mean value is expected to be near zero for signal events, 
and the \DeltaE\  resolution for signal events is about 18~\mev.
The beam-energy-substituted mass is defined by 
$\mes = \sqrt{({s}/{2} + {\vec{p}}_{i} \cdot  {\vec{p}}_{B} )^{2}/{E^{2}_{i}} - {{\vec{p}}_{B}}^{\,2}}$,
where $( {\vec{p}}_{i}, E_{i} )$ is the four-momentum of the initial-state 
\epem\ system and 
${\vec{p}}_{B}$ is the momentum of the $B$ candidate, 
both measured in the laboratory frame.
The \mes\ resolution for signal events is about 2.6~\mev.\ 
We retain candidates with  $|\DeltaE|<0.30$~\gev and 
$5.22<\mes<5.30$~\gev (this is referred to as the bounded \DeltaE-\mes\ plane).

Studies of off-resonance data and MC samples have shown that the largest background source
that passes the above cuts comes from continuum 
$e^+e^-\ra u\overline{u}/d\overline{d}/s\overline{s}/c\overline{c}$ events.
For this reason we choose to cut
on three variables that are commonly used to reject continuum background. 
The most powerful continuum-rejection cut 
is $|\cos\theta_{T}|<0.8,$ where $\cos\theta_{T}$ is the
cosine of the angle between the $B$ 
candidate thrust axis and the thrust axis of the remaining charged tracks and photons in the event.
The $|\cos\theta_{T}|$  distribution is fairly uniform for signal events, and is peaked near 1 for continuum
events. 
We also require $-5.0<{\cal F}<+1.0,$ where ${\cal F}$
is a Fisher discriminant \cite{bfish} based on zeroth and second momentum-weighted 
Legendre polynomial sums of the remaining tracks and photons 
$({\cal F}=0.5264 - 0.1882{\cal L}_0 + 0.9417{\cal L}_2),$ and 
$R_2<0.3,$ where $R_2$ is the ratio of second to zeroth Fox-Wolfram moments \cite{r2ref}.

After all the above cuts are applied, there are a small $(<1\% )$ number of events with more than
one $3K^0_s$ candidate per event. While the MC appears to properly model this
effect, we choose to simplify the analysis by using only the candidate with the
smallest $\Sigma (\dmpp^i)^2$ in each event, where the sum is
over the three $K^0_s$ candidates making up the $\b3ks$ candidate.

\section{MAXIMUM LIKELIHOOD FIT FOR YIELD}

We take all the combinations that pass the above cuts 
and extract the yield of signal $\b3ks$ events 
$(N_S)$ with an unbinned extended maximum likelihood (ML) fit, where the likelihood is:
\begin{equation}
{\mathcal L} = \exp{\left(-\sum_{i}N_{i}\right)}
\prod_{j=1}\left[\sum_{i}N_{i}{\mathcal P}_{ij}\right] ,
\end{equation}
the sum over $i$ corresponds to four categories of signal and background, 
the product over $j$ corresponds to 
the 508 $\b3ks$ candidates that pass all the requirements in the previous section, 
and ${\mathcal P}_{ij}$ is the probability
density function (PDF) for the $i$th category evaluated for the $j$th candidate.
The PDFs used for the four components of the fit are described in the next sections. 
For all except 
one category, 
the PDFs are the products of one-dimensional PDFs in \mes\ and \DeltaE :
$${\mathcal P}_{i}={\mathcal P}_{i}(m_{ES}){\mathcal P}_{i}(\Delta E).$$
The four categories whose yields $N_i$ are determined by the fit are the above-mentioned number
of signal events $N_S,$ 
the number of continuum BG events $N_{CBG},$ 
the number of $B\overline{B}$ events
in which the candidate comes from random combinations 
of tracks that may or many not be true $\k2pi$
decays (the ``non-peaking'' $\bab$ BG) $N_{BNOP},$ 
and events where the three $K^0_s$ candidates come from the
same $B$ decay but that $B$ decay is not a signal $\b3ks$ decay (the ``peaking'' BG) $N_{BPBG}.$
The default fit requires all $N_i\ge 0,$ but we loosen this requirement as a
systematic cross-check, and see that the signal yield changes by a small amount.

The dominant source of peaking BG in the part of the \DeltaE-\mes\ plane over which we do our fit is
$B\ra 3K^0_s\pi$ decays, where the $\pi^+$ or $\pi^0$ is missed.
Many of the $B$ branching fractions for decays that contribute to this background 
are known poorly or not at
all \cite{ref:pdg2002}, including the decays $B^0\ra 2K^0_sK^{* 0}$ and $B^+\ra 2K^0_sK^{* +},$
which should dominate the peaking BG in the bounded \DeltaE-\mes\ plane. 
Like the signal $\b3ks$ events, we remove the peaking BG from the generic $B\overline{B}$ MC
before we fit it to extract the parameters used to describe the non-peaking $\bab$ BG PDF. 
We generate large MC samples with only $\bab$ peaking BG, 
extract the peaking PDF from these samples,
and allow the $\bab$ peaking and non-peaking yields to float separately. 
One reason the $|\Delta E|$ cut is so wide $(<300$ MeV) on the bounded \DeltaE-\mes\ plane
is to allow a large enough region to fit the peaking BG. 
We do not attempt to extract a branching fraction for the peaking $\bab$ BG; 
our goal is to determine the size and shape of the peaking BG,  
so that its presence at an unexpectedly large level does not distort the shape of the 
non-peaking $\bab$ BG and lead to a systematic bias on $N_S.$ 
It turns out that the fit to the data requires very little peaking $or$ non-peaking $\bab$ BG, 
and the signal yield is changed negligibly 
if either or $both$ of these populations are fixed at zero. 

\subsection{SIGNAL EFFICIENCY AND PROBABILITY DENSITY FUNCTION FOR SIGNAL}

\begin{figure}[!htb]
\begin{center}
\includegraphics[height=7cm]{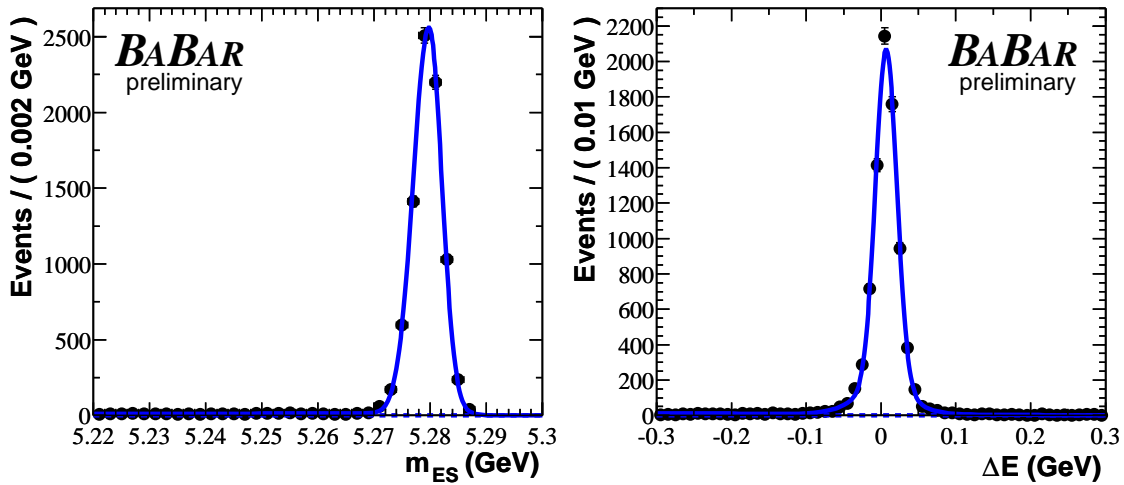}
\caption{Distributions of $m_{ES}$ and $\Delta E$  for selected candidates from 
MC simulated $\b3ks$ events. 
The curves overlaid on the MC data points are the signal PDFs determined from a unbinned fit.
The parameters from the fit are shown in Table 1.}
\label{fig:signal_pdfs}
\end{center}
\end{figure}

The signal PDF parameters are determined from ML fits to reconstructed and selected 
signal MC events from a sample of
148k generated $\b3ks$ decays. 
The signal
distribution in $\Delta E$ is well described by a double-Gaussian (one for the ``core,''
another for the broader ``tail''). The five parameters that describe 
this double-Gaussian PDF are shown in Table \ref{tab:fitparams}.
The signal PDF in $m_{ES}$ is well described by a bifurcated Gaussian.
The values of the three parameters determined from the fit to the signal MC 
are also shown in Table \ref{tab:fitparams}.
Histograms of the signal MC, with the derived PDFs shown as overlaid curves, are
shown in Fig.~\ref{fig:signal_pdfs}.
A fit to the signal MC with these parameters gives a signal yield 
of $N_S=8147\pm 91$ events, or a signal efficiency of $(5.50\pm 0.06)\% .$ 

\begin{table}[!htb]
\caption{Parameters used in the default ML fit. 
The parameters and their statistical errors are derived from MC studies described in the text.
The signal $\Delta E$ PDF is proportional to 
$f_{core}\exp ({-(\Delta E-\mu_{core})^2\over{2\sigma_{core}^2}})  
+(1-f_{core})\exp ({-(\Delta E-\mu_{tail})^2\over{2\sigma_{tail}^2}}),$ 
the signal $m_{ES}$ PDF is proportional to  
$\theta (\mu -m_{ES})\exp ({-(m_{ES}-\mu )^2\over{2\sigma_{left}^2}})  
+\theta (m_{ES}-\mu )\exp ({-(m_{ES}-\mu )^2\over{2\sigma_{right}^2}}),$ 
the continuum and non-peaking $\bab$ $m_{ES}$ PDFs are proportional to
$m_{ES}\sqrt{1-({m_{ES}\over{m_0}})^2}\exp (-\xi \lbrack 1-({m_{ES}\over{m_0}})^2\rbrack ).$}
\begin{center}
\begin{tabular}{|c|c|c|}
\hline  Fit component &  Parameter &   Value \\ \hline \hline    
        &   $\Delta E\ \mu_{core}$  &  $6.9\pm 0.2$ MeV \\
        &   $\Delta E\ \sigma_{core}$  &  $14.3\pm 0.2$ MeV \\
        &   $\Delta E\ \mu_{tail}$  &  $2\pm 1$ MeV \\
Signal  &   $\Delta E\ \sigma_{tail}$  &  $37\pm 1$ MeV \\
        &   $\Delta E\ f_{core}$  &  $(85\pm 1)\% $ \\
        &   $m_{ES}\ \mu$  &  $5279.8\pm 0.1$ MeV \\
      &     $m_{ES}\ \sigma_{left}$  &  $2.8\pm 0.1$ MeV \\
      &     $m_{ES}\ \sigma_{right}$  &  $2.3\pm 0.1$ MeV \\ \hline
Common BG &         $m_{ES}\ m_0$  &  5289.8 MeV \\ \hline
Continuum  &  $m_{ES}\ \xi$  & $-17\pm 8$ \\
BG  &        $\Delta E$ linear slope  & $-1.9\pm 0.3$ \\ \hline
Non-peaking & $m_{ES}\ \xi$  & $-47\pm 24$ \\
$\bab$ BG  &  $\Delta E$ linear slope  & $-3.2\pm 0.6$ \\ \hline
\end{tabular}
\end{center}
\label{tab:fitparams}
\end{table}

\subsection{PROBABILITY DENSITY FUNCTION FOR CONTINUUM}

\begin{figure}[!htb]
\begin{center}
\includegraphics[height=7cm]{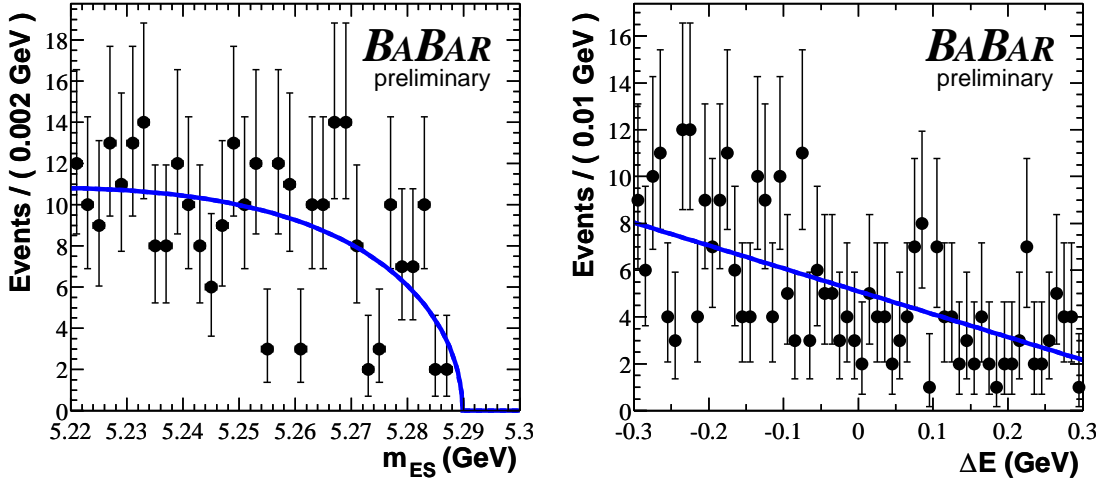}
\caption{Distributions of $m_{ES}$ and $\Delta E$  for selected candidates from 
$u\overline{u}/d\overline{d}/s\overline{s}/c\overline{c}$ MC simulated samples. 
The curves overlaid on the MC data points are the PDFs used to describe the continuum BG 
in the bounded $\Delta E-m_{ES}$ plane.}
\label{fig:cont_pdfs}
\end{center}
\end{figure}

Various studies suggest that a linear function
is sufficient to describe the continuum BG 
in $\Delta E,$ so we use that for the $\Delta E$ PDF. 
For $m_{ES}$ the standard PDF is the Argus function \cite{argus},
and this proves sufficient. 
There is one parameter (the slope) for the $\Delta E$ PDF and
two parameters $(m_0$ and $\xi )$ for the Argus function.
The parameters $m_0$ and $\xi$ are correlated, 
and if the $m_0$ parameter is set too low, 
there can be problems with the fits.
To avoid this problem, $m_0$ is fixed to 5.2898 GeV in the fit. 
It is, however,
allowed to float when systematic errors are evaluated.
Both the $\Delta E$ slope and $\xi$ are floated in a fit to a continuum 
MC sample corresponding to 77 ${\rm fb}^{-1}$ of integrated luminosity. 
The projections of the MC events and the continuum BG PDFs are shown in
Fig.~\ref{fig:cont_pdfs}. 
The values of the parameters used in these PDFs and in subsequent
fits are shown in Table \ref{tab:fitparams}.

\subsection{PROBABILITY DENSITY FUNCTION FOR NON-PEAKING $B\overline{B}$ BACKGROUND}

The non-peaking $\bab$ BG has a similar source as the continuum BG 
(random combinations of real and fake $K^0_s$ mesons) 
and so is expected to have a similar functional form.
We used the same description (linear function
for $\Delta E$ and Argus function for $m_{ES}$) as used for the continuum BG 
and fit them to $B\overline{B}$ MC samples with any candidates identified 
as signal or peaking $\bab$ BG removed. 
We used a $m_0$ parameter in common with the continuum BG, 
but the $\Delta E$ slope and Argus $\xi$ parameter 
are allowed to float to different values from those used for continuum BG. 
The values determined by the fit to the non-peaking $\bab$ MC samples
are also shown in Table \ref{tab:fitparams}.

\subsection{PROBABILITY DENSITY FUNCTION FOR PEAKING $B\overline{B}$ BACKGROUND}

The distributions for the peaking $\bab$ BGs have a very different functional form from those
for the continuum and non-peaking $\bab$ BGs.
We have parametrized the peaking $\bab$ BG as was done in Ref. \cite{denisbr}, 
by taking the two-dimensional (2D) histogram of all candidates (which pass the analysis cuts)
from specially generated $B^0\ra 2K^0_sK^{* 0}$ and $B^+\ra 2K^0_sK^{* +}$ 
MC samples and using this 2D histogram as the PDF.

\begin{figure}[!htb]
\begin{center}
\includegraphics[height=7cm]{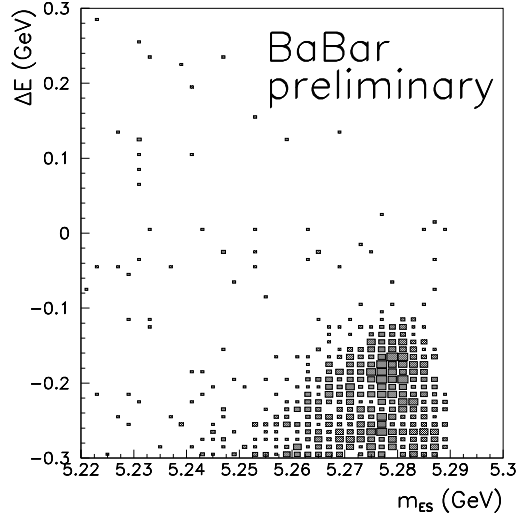}
\caption{The two-dimensional histogram of $\Delta E$ vs.~$m_{ES}$ used as the PDF 
for the peaking $\bab$ BG.}
\label{fig:2d_peaking}
\end{center}
\end{figure}

The 2D histogram PDF, with the same binning used in the default fit, is shown
in Fig.~\ref{fig:2d_peaking}. By using a 2D histogram for the PDF, a large correlation
between $\Delta E$ and $m_{ES},$ not present for the continuum or non-peaking $\bab$
BGs, is properly taken into account. Systematic errors due to the binned nature
of this PDF are discussed later.

\section{FIT FOR YIELD}

We use the PDFs described in the preceding sections and the parameters derived 
from fits to MC samples and listed in Table \ref{tab:fitparams} for our
default fit to the on-resonance data sample.
The values for the populations of the various components of the fit are 
$$(N_{S},N_{CBG},N_{BNOP},N_{BPBG})=(71\pm 9,
428^{+23}_{-29},
0^{+26}_{-0},
9\pm 8).$$

\begin{figure}[!htb]
\begin{center}
\includegraphics[height=7cm]{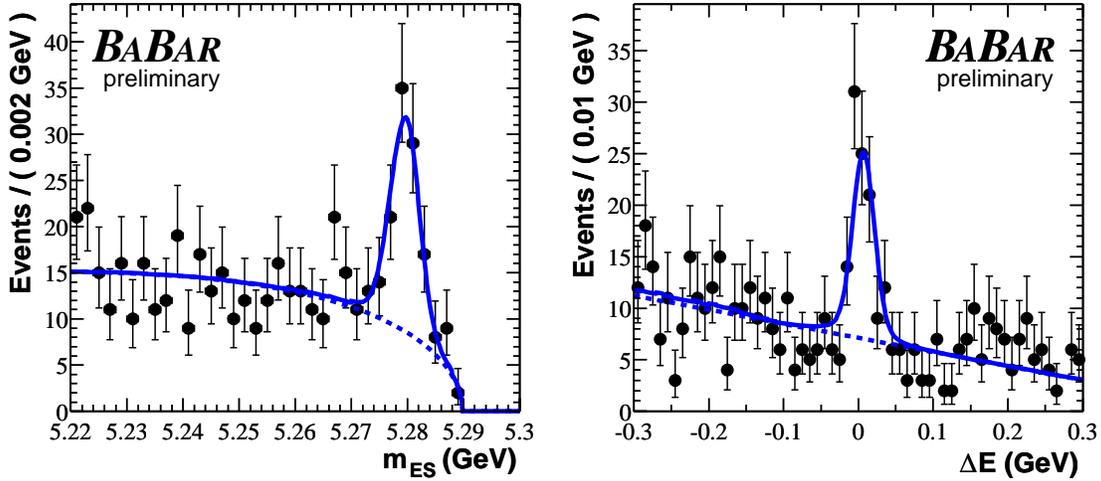}
\caption{Distributions of $m_{ES}$ and $\Delta E$ for 
selected candidates from the on-resonance data sample. 
The solid curve overlaid on the data corresponds to the sum of all PDFs, 
with their parameters (Table \ref{tab:fitparams}) 
and the signal and background fractions returned by the fit. 
The dashed curve is the contribution from the continuum BG.}
\label{fig:dataon_run1234_default_fit}
\end{center}
\end{figure}

The fit requires no $\bab$ non-peaking BG, but does allow for a small but
not significant amount of $\bab$ peaking BG. The projections of the data and the fits
on the $m_{ES}$ and $\Delta E$ axes are shown in Fig.~\ref{fig:dataon_run1234_default_fit}. 
The small contribution of the $\bab$ peaking BG is the non-overlap of the 
solid line (all components of the fit) and the dashed line (the continuum BG component
of the fit) in the region $\Delta E <-0.10$ GeV in the $\Delta E$ plot. 

\section{SYSTEMATIC STUDIES}
\subsection{CUT-AND-COUNT ANALYSIS}
\label{sec:Systematics}
We use a simple ``cut-and-count'' analysis to cross-check the results
of the maximum likelihood fit, and to study several
sources of systematic uncertainty. 
For the cut-and-count
analysis we define a signal region centered on the expected signal
in \mes\ and \DeltaE . The signal region is defined by $5.2704<m_{ES}<5.2884$ GeV
and $-40<\Delta E<+40$ MeV. We define two \DeltaE\ sidebands
with the same \mes\ cut but with $-300<\Delta E<-100$ MeV and
$+100<\Delta E<+300$ MeV. The sum of the number of entries in the
sidebands (57) scaled by the ratio of the area in the signal region to the area in the sidebands
(0.2) gives an estimate of the number of BG events in the signal region $(11\pm 2),$
where 78 events are observed, for a signal yield of $67\pm 9$ events.
The signal efficiency is slightly different between the
cut-and-count analysis and the ML fit. 
The MC efficiency-corrected yield is
$1295^{+170}_{-158}$ events for the ML fit 
and $1258\pm 169$ events for the cut-and-count analysis. 
Given the different methods for estimating signal and background in the ML fit and
the cut-and-count analysis, they are in reasonable agreement. 

\subsection{SIGNAL EFFICIENCY VARIATION ACROSS DALITZ PLOT}

The cut-and-count analysis allows one to
take all the entries in the signal region and plot the $m_{2K^0_s}$ distributions
for these candidates, and 
compare them to
those predicted by the reconstructed signal MC, which was
generated with the assumption of 
non-resonance phase-space (uniform population of the Dalitz plot at generation). 
The distributions of the reconstructed $m_{2K^0_s}$ masses
are consistent with (reconstructed) three-body
phase space, but with such a small number of events in the data, we cannot
rule out resonance production at the level of a few events per resonance in
our data sample, or other small deviations from phase-space.

\begin{figure}[!htb]
\begin{center}
\includegraphics[height=7cm]{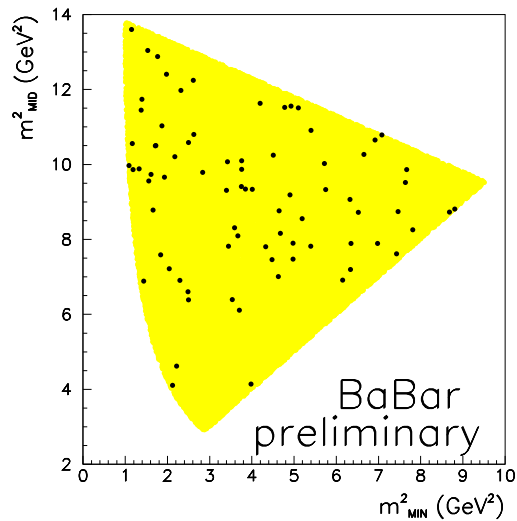}
\caption{The folded Dalitz plot for the 78 candidates (black points) that pass the cut-and-count
analysis and end up in the signal region. The signal/background for this selection
is $\sim 5.8.$ The yellow-shaded area is the physically-allowed region.}
\label{fig:conf_fig6}
\end{center}
\end{figure}

This is important because the efficiency calculated in
different regions of the Dalitz plot is not uniform. It
varies by more than a factor of two, mostly due to low
efficiency for reconstructing $K^0_s$ near some
edges of the Dalitz plot. If the parent distribution for the data
is the same as the MC parent distribution, this is not a
problem, but since we do not know this, we need to assign
a systematic error for the possibility that it is not. 

We do this by dividing the ``folded'' Dalitz plot 
(one of the unique sextants of the $3K^0_s$ Dalitz plot) 
into 21 bins and calculating an efficiency
for the cut-and-count analysis for each bin individually. 
The folded Dalitz plot we use is shown as the shaded area
in Fig.~\ref{fig:conf_fig6}; the points are the 78 candidates
in the signal region of the cut-and-count analysis for the data.
This folded Dalitz is achieved by ordering the three unique $m(2K^0_s)$ combinations
for each $\b3ks$ candidate $m_{MAX}(2K^0_s)>m_{MID}(2K^0_s)>m_{MIN}(2K^0_s)$
and plotting $m_{MID}(2K^0_s)$ vs.~$m_{MIN}(2K^0_s).$  

The entries in the signal region are then individually corrected for
efficiency depending on what bin they populate, and the
yield calculated this way is compared to the yield when
all events are given the same (average) efficiency.
The yield differs by 4.2\%\  between the two ways of calculating the efficiency; 
we take this as the systematic error
estimate due to a nonuniform population of the Dalitz plot.

\subsection{$K^0_s$ RECONSTRUCTION EFFICIENCY}

There is a small but well-measured disagreement between the
$\k2pi$ reconstruction efficiency in the data and the one reported by
the full detector MC. We correct the efficiency and
calculate a systematic error on how well we
know the $K^0_s$ reconstruction efficiency. 
The efficiency in the MC simulation and data is measured as a function of 
$K^0_s$ transverse decay radius $(r_{DEC}),$
transverse momentum and polar angle in the \babar\ detector, 
and also for periods with different detector running conditions.
A correction is calculated for each
of the three $K^0_s$ candidates in a reconstructed MC event, and the
product of the three correction factors (taken as the $B^0$ candidate
correction factor) is averaged over all selected events in the signal MC sample.
We do this for several
different sets of 
measured efficiencies,
each produced with different 
$K^0_s$ selection criteria.
The corrections evaluated for different selection criteria are
consistent, and the average correction factor 
is $\varepsilon_{data}/\varepsilon_{MC}=0.950\pm 0.014.$

The error includes a statistical error for the tables
used to calculate the correction, a per-charged-track systematic error, and a per-$K^0_s$
systematic error. The quadrature sum of all these error estimates is $10.1\% .$
This is the dominant source of systematic error for this measurement.

\subsection{SIGNAL PARAMETRIZATION}

We allow each of the eight parameters in the signal PDF to float in the fit, one at a time.
The quadrature sum of the change in signal yield from these eight variations is $4.3\% ,$ and
we take this as the systematic error estimate on our signal parametrization. Since many of the
signal parameters are correlated, we also perform a fit where six of the signal parameters are
free; only the \DeltaE\ tail-Gaussian mean and width are fixed to the values derived from
the MC. The change in $N_S$ from letting all these parameters float together is $2.4\% .$ 
This is a variant of the above (larger) estimate that allows correlations between parameters
to be taken into account, but we will use the larger estimate as a more conservative estimate.

\subsection{BACKGROUND PARAMETRIZATION}

As with the signal parameters, we allow each of the five BG PDF parameters to float in the
fit, one at a time. For the binned-histogram peaking $\bab$ PDF, we increase and decrease the bin
size by a factor of two and allow different levels of smoothing of the histogram. We take
the largest of these variations as the systematic error due the peaking $\bab$ PDF, and add it in quadrature
with the changes in signal yield from letting the BG fit parameters float. The fractional systematic
error estimated from all these variations is $0.8\% .$ While the fit will not support all five
BG parameters floating at once, we take the two parameters with the largest correlation (the $\Delta E$
slopes for the continuum and non-peaking $\bab$ BGs) and let them float together. The signal yield
changes by $0.4\% ,$ less than the quadrature sum of the two parameters allowed to float separately.
While the populations of the BG categories change noticeably with all these parameter variations,
the total BG yield and the signal yield are quite stable.

\subsection{FIT VALIDATION}

We perform studies with parametrized MC simulations in which 
many samples of the same size (and category populations) as the data
are generated from the PDFs.  
We also perform MC studies in which the background events are generated from
the PDFs but the signal samples were extracted from the full-detector MC
sample and fit along with the background samples. 
The means and uncertainties for the yields are all
consistent with expectations, and no significant corrections for biases or systematic error
contributions are required. 
For 2000 fitted toy MC samples, 48\%\ have a larger value of $-\ln{\cal L}$ than that for the fit
to the data.

There is a 15.9 ${\rm fb}^{-1}$ sample of data 
taken at CM energies just below the $\Upsilon (4S)$ resonance (the
off-resonance data sample), which contains no $B$-meson decays. This data is corrected
for a shift in the \mes\ endpoint due to different beam energies. The data is
subjected to the same selection cuts and the same ML fit as the on-resonance data. The signal
yield from this fit is consistent with zero events.

\subsection{OTHER FIT VARIATIONS}

As a measure of the sensitivity to background parameterizations, 
we remove the three BG categories one at a time in our default ML fit. 
With either
(or both) of the $\bab$ BG PDFs removed, the yield changes by very little $(<0.3\% ).$
With the continuum BG PDF removed, the signal yield changed noticeably $(-5.1\% ),$
but the likelihood of the no-continuum-BG fit is much worse than that for the default fit.
That is, neither of the $\bab$ BGs (or their combination) does a particularly
good job of describing the continuum BG, which dominates the bounded \DeltaE -\mes\ plane
away from the signal region. 

We remove the restriction that each of the four yields be greater than zero and refit. 
The signal yield changes
by $+1.3\% ,$ and the likelihood of this fit is only slightly better.

We add (separately) quadratic terms to the continuum and non-peaking $\bab$ $\Delta E$ PDFs
and let them float in the fit; the signal yield changes by $-2.2\% $ and $-0.6\% .$ 
We include these variations in the systematic uncertainty 
along with the other systematic errors estimated from floating the
background parameters above. 

\subsection{CANDIDATE SELECTION CRITERIA}

While the candidate selection cuts listed in Section 3 are quite standard 
and we expect the MC to reproduce them, we 
estimate a systematic error on each one to account for the fact that the MC might
not exactly reproduce the data. 
Where possible, each cut is removed in turn 
and the change in yield for the data is compared with that for the
MC. For the few cuts that are significantly correlated, both cuts are
removed at the same time. The cuts are also tightened by reasonable
amounts and the change in yield in the data and MC are compared.
Various other studies are performed on control channels such as 
$B^0\rightarrow D^{* -}\pi^+,$ 
$D^{* -}\ra\pi^-_s\overline{D}^0,$ 
$\overline{D}^0\ra K^0_s\pi^+\pi^-,$ which has a similar topology
and a large enough branching fraction so that it can be reconstructed
with minimal cuts. The quadrature sum of the
systematic error estimated from a variation for 
each cut is $5.0\% ,$ where the dominant contributions are from the
$\dmpp$ cut $(3.0\% ),$ the $\cos\theta_T$ cut $(2.5\% ),$ and the
$R_2$ cut $(2.1\% ).$

\subsection{SYSTEMATIC ERROR SUMMARY}

There are two other small systematic errors shown in Table \ref{tab:sysest} not discussed
above: the statistical error on the MC used to derive the signal efficiency estimate,
and the error on the total number of $B\overline{B}$ pairs in our data sample.
With these added in quadrature with the systematic error estimates described above, the total
(fractional) systematic error is $13.1\% .$

\begin{table}[!htb]
\caption{Summary of fractional systematic uncertainties.}
\begin{center}
\begin{tabular}{|c|c|c|}
\hline Source & Estimated from & Percent Error  \\ \hline\hline
$K^0_s$ efficiency & detector studies & 10.1\%         \\
BG Parametrization & vary in fit & 2.4\%       \\ 
Signal Parametrization & vary in fit & 4.3\%      \\ 
Candidate Selection Cuts & cut variations, studies & 5.0\%       \\ 
Efficiency variation & cut-and-count analysis & 4.2\%       \\ 
Signal efficiency & MC statistics & 1.3\%       \\ 
$\bab$ counting &  & 1.1\%       \\ \hline\hline
Total &   & 13.1\%  \\
\hline
\end{tabular}
\vspace{0.4cm}
\end{center}
\label{tab:sysest}
\end{table}

\section{PHYSICS RESULTS}
\label{sec:Physics}

The branching fraction is calculated from the relationship ${\cal B}=N_S/(\varepsilon N_{B\overline{B}}),$
where $N_S=71\pm 9$ is the signal yield from the fit, 
and $\varepsilon$ is the
product of $\varepsilon_{MC}=5.50\% $ derived from the signal MC 
and $\varepsilon_{data}/\varepsilon_{MC}=95.0\% ,$
derived from the $K^0_s$ efficiency studies. 
These combine for an efficiency-corrected signal yield
of $1363^{+179}_{-167}.$ 
The data set corresponds to $211\times 10^{6}$ $B^0$ and $\ab0$ decays,
and we calculate ${\cal B}=(6.5\pm 0.8)\times 10^{-6}$ 
(statistical error only). 
We assume
that the rate for $B^0\ab0$ and
$B^+B^-$ production in $\Upsilon (4S)$ decays is equal. 
The errors on all quantities except for the signal
yield are included in the systematic error estimate.
If we fix the signal yield to zero in our ML fit, the difference between the $-\ln{\cal L}$
of this fit and the default fit gives a statistical significance for our observation
of $15.6\sigma .$

The sum of the systematic error estimates is given in Table \ref{tab:sysest}. 
This results in a measurement of ${\cal B}=(6.5\pm 0.8\pm 0.8)\times 10^{-6},$
where the second error is the systematic error estimate. One sigma of the total systematic
error estimate (not just the ones that pertain to the signal yield) 
corresponds to 9.3 (efficiency-uncorrected) events. When the signal yield is fixed
to 9.3 and the data is refit, the change in $-\ln{\cal L}$ from the default fit 
corresponds to a significance of $10.9 \sigma .$
 
We note that this measurement is consistent with the previous Belle measurement, but
by itself it is more than $2\sigma$ above the prediction made using ${\cal B}(B^+\ra K^+K^-K^+)$
and the assumption of penguin dominance in $B\ra KKK.$ 
However, if this difference is confirmed with more data, 
it may just be evidence of resonant intermediate states
occurring at different rates in $\b3ks$ and $(\phi -$removed) $B^+\ra K^+K^-K^+.$

While we have examined the 
folded Dalitz plot for 
the cut-and-count analysis and
see nothing that looks like a narrow resonance (broad resonances cannot be ruled out
given the size of the data sample), decays like $B^0\ra \chi_{c0}K^0$ may be
present and are clearly not part of the relationship between 
$\b3ks$ and $B^+\ra K^+K^-K^+.$ 
To estimate the amount of this type of decay
in our sample, we reject $2K^0_s$ masses within $\pm 50$ MeV (about $3\sigma$ of our resolution)
of the $\chi_{c0}$ and $\chi_{c2}$ masses. We apply this rejection to the data and the MC
samples in our cut-and-count analysis and, while a few entries are removed from the data,
proportionally slightly more were removed from the non-resonant phase-space generated 
MC, so the efficiency-corrected yield
goes $up$ slightly (consistent with no change) when the $\chi_{c}$ bands are excluded.
On the basis of this, we cannot claim we observe any contribution to the $\b3ks$ signal
from $B^0\ra \chi_{c}K^0,$ and so we leave the inclusive measurement uncorrected.

\section{SUMMARY}
\label{sec:Summary}
From a sample of $211\times 10^6$ $B\overline{B}$ decays recorded with the \babar\ detector,
we observe a signal of $71\pm 9$ $\b3ks$ decays, and with these measure
a branching fraction ${\cal B}(\b3ks )=(6.5\pm 0.8 \pm 0.8 ) \times 10^{-6}.$
This result is preliminary.

\section{ACKNOWLEDGMENTS}
\label{sec:Acknowledgments}

We are grateful for the 
extraordinary contributions of our \pep2\ colleagues in
achieving the excellent luminosity and machine conditions
that have made this work possible.
The success of this project also relies critically on the 
expertise and dedication of the computing organizations that 
support \babar.
The collaborating institutions wish to thank 
SLAC for its support and the kind hospitality extended to them. 
This work is supported by the
US Department of Energy
and National Science Foundation, the
Natural Sciences and Engineering Research Council (Canada),
Institute of High Energy Physics (China), the
Commissariat \`a l'Energie Atomique and
Institut National de Physique Nucl\'eaire et de Physique des Particules
(France), the
Bundesministerium f\"ur Bildung und Forschung and
Deutsche Forschungsgemeinschaft
(Germany), the
Istituto Nazionale di Fisica Nucleare (Italy),
the Foundation for Fundamental Research on Matter (The Netherlands),
the Research Council of Norway, the
Ministry of Science and Technology of the Russian Federation, and the
Particle Physics and Astronomy Research Council (United Kingdom). 
Individuals have received support from 
CONACyT (Mexico),
the A. P. Sloan Foundation, 
the Research Corporation,
and the Alexander von Humboldt Foundation.

\end{document}